\documentclass[aps,10pt,twocolumn,showpacs,nofootinbib,prd]{revtex4-1}%   
\usepackage[colorlinks,citecolor=DarkGreen,linkcolor=DarkRed,urlcolor=DarkBlue]{hyperref}
\usepackage{amsmath}
\usepackage{graphicx}
\usepackage{slashed}
\usepackage[svgnames]{xcolor}

\begin{document}

\title{Magnetic field dependence of nucleon parameters from QCD sum rules}

\author{C. A. Dominguez$^1$}
\author{Luis A. Hern\'andez$^{1,2,3}$}
\author{Marcelo Loewe$^{4,5,1}$}
\author{Cristian Villavicencio$^{6}$}
\email[E-mail: ]{cvillavicencio@ubiobio.cl}
\author{R. Zamora$^{7,8}$}

\affiliation{$^1$Centre for Theoretical \& Mathematical Physics, and Department of Physics, University of Cape Town, Rondebosch 7700, South Africa;}
\affiliation{$^2$Instituto de Ciencias Nucleares, Universidad Nacional Aut\'onoma de M\'exico, Apartado Postal 70-543, CdMx 04510, Mexico;}
\affiliation{$^3$Facultad de Ciencias de la Educaci\'on, Universidad Aut\'onoma de Tlaxcala, Tlaxcala, 90000, Mexico;}
\affiliation{$^4$Instituto de F\'isica, Pontificia Universidad Cat\'olica de Chile,
Casilla 306, Santiago, Chile;}
\affiliation{$^5$Centro Cient\'ifico Tecnol\'ogico de Valparaíso-CCTVAL,
Universidad T\'ecnica Federico Santa Mar\'ia, Casilla 110-V, Valpara\'iso, Chile;}
\affiliation{$^6$Centro de Ciencias Exactas \& Departamento de Ciencias B\'asicas, Facultad de Ciencias, Universidad del B\'io-B\'io,
Casilla 447, Chill\'an, Chile;}
\affiliation{$^7$Instituto de Ciencias B\'asicas, Universidad Diego Portales, Casilla 298-V, Santiago, Chile;}
\affiliation{$^8$Centro de Investigaci\'on y Desarrollo en Ciencias Aeroespaciales (CIDCA), Fuerza A\'erea de Chile,  Santiago, Chile.}

% \affiliation{}

%\date{\today}

\begin{abstract}
Finite energy QCD sum rules involving nucleon current correlators are used to determine several QCD and hadronic parameters in the presence of an external, uniform, large magnetic field.
The continuum hadronic threshold $s_0$, nucleon mass $m_N$, current-nucleon coupling $\lambda_N$, transverse velocity $v_\perp$, the spin polarization  condensate $\langle\bar q\sigma_{12} q\rangle$, and the magnetic susceptibility of the quark condensate $\chi_q$, are obtained for the case of protons and neutrons.
Due to the magnetic field, and charge asymmetry of light quarks {\it up} and {\it down}, all the obtained quantities  evolve differently with the magnetic field, for each nucleon or quark flavor. With this approach it is possible to obtain the evolution of the above parameters up to a magnetic field strength $eB< 1.4$\,GeV$^2$. 
\end{abstract}

\maketitle

\section{Introduction}

The influence of external, strong magnetic fields on hadronic and Quantum Chromodynamics (QCD) properties is an active research field. 
There are two important physical scenarios  involving  these extreme magnetic fields, i.e. peripheral heavy ion collisions, and compact stars such as magnetars.
Several methods have been employed in order to extract the magnetic evolution of various quantities, e.g.  masses, coupling constants, and QCD vacuum condensates. These are, among others, lattice QCD, linear sigma model, Walecka model, Nambu--Jona-Lasinio model, functional renormalization group, and QCD sum-rules. 

Lack of experimental data makes it imperative to explore  different perspectives for a better understanding of such systems.
There are a few open questions in this field, e.g. the magnetic behavior of vacuum condensates. 
However, there is a consensus with respect to the existence of magnetic catalysis of the chiral condensate at zero temperature \cite{Kharzeev:2013jha}. 
For this reason, the chiral condensate at finite magnetic field can be considered as a reliable input in several analyses. 
The magnetic behavior of other condensates, e.g. the gluon condensate, is still not well established. 
However, there seems to be a reasonable consensus within a certain region of the magnetic field strength \cite{Dominguez:2018njv,DElia:2015eey}.
In the case of new condensates appearing in the presence of an external magnetic field, the main object is the polarization tensor condensate $\langle\bar q\sigma_{\mu\nu} q\rangle$.
All approaches agree on the linear behavior of this object for weak magnetic fields \cite{Kharzeev:2013jha,Frasca:2011zn}.

The baryon sector, in particular the nucleon case, in magnetic fields has been less explored. 
As a result, there are several pending issues. 
For instance, the magnetic behavior of the nucleon mass is unclear. 
In fact, different approaches lead to different  behavior. 
Some models lead to an increasing mass with increasing field, others lead to the opposite behavior, and still others suggest no significant magnetic field dependence \cite{Yue:2008tp,Haber:2014ula,Taya:2014nha,Mukherjee:2018ebw,Endrodi:2019whh,Coppola:2020mon}.

QCD sum rules have been used to study magnetic field effects in baryonic systems. Early work was focused on the determination of the nucleon magnetic moment  in the presence of new condensates \cite{Ioffe:1983ju}. This procedure was later extended  to the full baryonic octet \cite{Wang:2008vg}. These approaches considered only linear magnetic field dependences, and included the new condensate.
We consider here Finite Energy QCD sum rules (FESR) in an external magnetic field to obtain the behaviour of various QCD and hadronic parameters. 
The analysis is not restricted to linear magnetic field dependences. 
These sum rules allow for the  extraction of information about the hadronic continuum threshold $s_0$, the current-nucleon coupling  $\lambda_N$, the transverse velocity of nucleons, the polarization tensor  of the condensate, the magnetic susceptibility of the quark condensate, and the nucleon masses.
For this purpose, techniques used in previous work are implemented \cite{Ayala:2015qwa,Dominguez:2018njv,Villavicencio:2020fcz}.\\

This paper is organized as follows. 
In Section\,\ref{sect_FESR-vac} we present a brief introduction to FESR, and their application to the vacuum correlation of nucleonic interpolating currents. 
Section\,\ref{sec_magnetic} is devoted to  the introduction of an external magnetic field, specifying in detail the changes in propagators, correlator structures, and the corresponding modifications induced in the FESR.
In section\,\ref{sec_results} we present the set of QCD sum rules together with the numerical analysis. 
The article ends with the conclusions in section\,\ref{sec_conclusions}.

\section{Vacuum FESR }
\label{sect_FESR-vac}

\subsection{Brief description of FESR}

The FESR involve current-current correlators in momentum space $\Pi(p^2)$ multiplied by some analytic kernel (for a review see e.g.\cite{Dominguez:2018zzi}). 
These correlators are integrated along a closed contour consisting of a circular path in the complex plane $s=p^2$, skipping the singularities and branch cut that lie along the real positive $s$-plane (the \emph{pac-man} contour) \cite{Ayala:2015qwa,Dominguez:2018njv,Villavicencio:2020fcz}.  
From Cauchy's theorem the contour integral can be expressed as 
\begin{equation}
\frac{1}{\pi} \int_{0}^{s_0} \!\!ds\, K(s)\, {\mbox{Im}}\, \Pi(s)|_\text{\tiny{Had}} = 
  \frac{-1}{2 \pi i} \oint_\text{\tiny{$s_0$}}\!\! ds\,K(s)\,\Pi(s)|_\text{\tiny{QCD}},  \label{FESR}
\end{equation}
where $K(s)$ is an analytic kernel, usually a power of $s$, and the integration path on the right hand side is a circle of radius $s_0$. 
This expression encapsulates the notion of quark-hadron duality, with hadronic information along the positive real axis being related to QCD on the circle.  One of the advantages of FESR with respect to other sum rules  is that they provide a projection of vacuum condensates of a given mass dimension for each power of $s$. 

\subsection{Nucleon FESR}

The issue of the correlator involving two baryonic currents was pioneered in  \cite{Ioffe:1981kw}. 
Subsequently, a different baryonic current was considered in 
\cite{Chung:1981wm,Chung:1981cc,Chung:1982rd,Chung:1984gr}. 
However, this latter version implies that the correlator receives no contribution from  the lowest order chiral symmetry breaking operators. 
We shall use here the proton interpolating current of \cite{Ioffe:1981kw}
\begin{equation}
    \eta _{N} (x) = \varepsilon ^{abc}[(u^a)^T (x)\; C\gamma _{\mu}u^b (x)]\gamma ^{\mu} \gamma _5 \,d^c (x), \label{eta}
\end{equation}
where $C$ is the charge conjugation matrix given by $C = i \gamma_0\gamma_2$.  
For the neutron case one exchanges the quark flavors, i.e. $u\leftrightarrow d$ . 
In the hadronic sector $\eta (x)$ is defined  as
\begin{equation}
\langle 0 | \eta _{N} (x) |N(p, s) \rangle = \lambda _{N} u(p,s)e^{ip\cdot x}, \label{Lambda}
\end{equation}
with $\lambda _{N}$, the current-nucleon coupling, a phenomenological parameter \textit{a priori} unknown,  and $u(p,s)$ is the nucleon spinor. 
The interpolating function can then  be expressed in terms of the nucleon field as $\eta_N(x) =\lambda_N\Psi_N(x)$, and the two-point function is
\begin{equation}
   \Pi(q) = i \int d^4 x  \; e^{iq x}  \langle 0| T\,\eta (x) \bar{\eta } (0)| 0\rangle.  \label{Pi}
\end{equation}

In  the vacuum, this correlation function involves two independent structures 
\begin{equation}
\Pi (q)  = \slashed{q} \, \Pi _{1}(q^2) + \Pi _{2}(q^2). \label{Pi12}
\end{equation}
This correlator obeys the Operator Product Expansion (OPE) in the QCD sector. 
This includes a perturbative contribution associated to a two-loop Feynman diagram, in absence of radiative corrections, as well as a power series of non-perturbative condensates, led by the chiral-quark and the gluon condensates. 

In the chiral limit, neglecting radiative corrections, the  structures $\Pi _{1}$ and $\Pi _{2}$ in Eq.(\ref{Pi12}) of the correlator in the QCD sector are \cite{Ioffe:1981kw,Reinders:1984sr,Sadovnikova:2005ye,Nasrallah:2013ywh}
\begin{align}
\Pi_1(s) &= -\frac{1}{64\pi^4}\, s^2\ln(-s/\nu^2)
    -\frac{1}{32\pi^3}\langle\alpha_s G^2\rangle\ln(-s/\nu^2)
        \nonumber\\&\qquad
    -\frac{2}{3}\frac{\langle\bar qq\bar qq\rangle}{s}
    +C_8\frac{\langle{\cal O}_8\rangle}{s^2}
    +C_{10}\frac{\langle{\cal O}_{10}\rangle}{s^3}+\dots,\label{Pi1}\\
&\nonumber\\
\Pi_2(s) &= \frac{1}{4\pi^2}\langle\bar qq\rangle s\ln(-s/\nu^2)
    -\frac{1}{12\pi}\frac{\langle\alpha_s G^2\,\bar qq\rangle}{s}\
        \nonumber\\&\qquad
    +C'_9\frac{\langle{\cal O}_9\rangle}{s^2}
    +C'_{11}\frac{\langle{\cal O}_{11}\rangle}{s^3}+\dots,  \label{Pi2}
\end{align}
where $\nu $ is the  dimensional regularization scale, and the correlator components  in the hadronic sector are
\begin{align}
\Pi_1(s)    &=  \frac{-\lambda_N^2}{s-m_N^2},
& \Pi_2(s)    &= \frac{-\lambda_N^2  m_N}{s-m_N^2}. \label{Pi12HAD}
\end{align}

Integrating the two components of the correlator in Eq.\,(\ref{FESR}), and using the FESR kernel $K=1$ there follow two equations relating the hadronic to the QCD sector
%Eq.9-10
\begin{align}
\lambda_N^2       &=  \frac{s_0^3}{192\pi^4}
    + \frac{s_0}{32\pi^3}\langle\alpha_s G^2\rangle 
    +\frac{2}{3}\langle\bar qq\bar qq\rangle, 
        \label{eq_FESR-Pi1} \\ 
\lambda_N^2m_N  &=  -\frac{s_0^2 }{8\pi^2}\langle\bar qq\rangle 
    + \frac{1}{12\pi}\langle\alpha_s G^2\,\bar qq\rangle.
        \label{eq_FESR-Pi2}
\end{align}
Invoking vacuum dominance for the four-quark condensate in Eq.\,(\ref{eq_FESR-Pi1}) and for the mixed quark and gluon condensate in Eq.\,(\ref{eq_FESR-Pi2}), a nucleon mass $m_N=0.94$\,GeV, the quark condensate $\langle \bar qq\rangle =-$(0.24\,GeV)$^3$, and the \emph{standard} value of the gluon condensate $\langle \alpha_s G^2\rangle = 0.037$\,GeV$^4$ \cite{Dominguez:2018zzi}, we obtain the values of the current-nucleon coupling $\lambda_N = 0.017$~GeV$^3$ and the hadronic continuum threshold $s_0=1.26$~GeV$^2$. 

The next step  is the evaluation of the FESR in the presence of a constant and uniform magnetic filed.

\section{Magnetic field effects}
\label{sec_magnetic}

\subsection{Propagators}

The fermion propagator in the presence of a constant and uniform magnetic field can be written as a power series in $qB$~\cite{Mukherjee:2018ebw}. 
Taking into account the anomalous magnetic moment term ${\cal L}_\text{\tiny anom}=-\frac{1}{2}\kappa\sigma_{\mu\nu}F^{\mu\nu}$ the fermion propagator becomes
%Eq.11
\begin{align}
S(p)= &\;\frac{i(\slashed{p}+m)}{p^2-m^2+i\epsilon}
-(qB)\frac{i\sigma_{12}(\slashed{p}_\parallel+m)}{(p^2-m^2+i\epsilon)^2}
\nonumber\\&
+2i(qB)^2\frac{(\slashed{p}_\parallel+m)\left[p_\perp^2-\slashed{p}_\perp(\slashed{p}_\parallel-m)\right]}{(p^2-m^2+i\epsilon)^4}
\nonumber\\&
-(\kappa B)\frac{i(\slashed{p}+m)\sigma_{12}(\slashed{p}+m)}{(p^2-m^2+i\epsilon)^2}
 +\dots,
\label{eq_SB}
\end{align}
where only relevant terms are considered, and  $\sigma_{\mu\nu}\equiv \frac{i}{2}[\gamma_\mu,\gamma_\nu]$ is the Dirac anti-symmetric tensor.\footnote{
	Notice that in the literature $\sigma_{\mu\nu}$ is often defined with the opposite sign.
	This will produce a change in the sign of the condensates $\langle \bar q\,\sigma_{12}\, q\rangle$.}
The mass $m$, charge $q$ and anomalous magnetic moment $\kappa$ correspond to the respective particles. 
The charges are $e_p = e$ for protons, $e_n=0$ for neutrons, $e_u=2e/3$ for the $u$-quark and $e_d=-e/3$ for the $d$-quark.
The magnetic anomalous moments are  $\kappa_p=1.79\mu_N$ for protons, $\kappa_n=-1.91\mu_N$ for neutron \cite{Patrignani:2016xqp} and $\kappa=0$ for quarks, with $\mu_N=e/2m_N$ being the nucleon Bohr magneton.

In the case of nucleons, the spatial asymmetry generated by the presence of the external magnetic field will be reflected in the effective nucleon propagator by considering a \emph{transverse velocity} term. 
The external momentum in the nucleon propagator is written as $p = p_\parallel +v_\perp p_\perp$.
The particle velocity is a medium  effect, often considered in pionic dynamics 
\cite{Kamikado:2013pya,Pisarski:1996mt,Son:1999cm}, and smaller than the speed of light. In the case of massless particles $v_\perp$ is simply the transverse velocity of the particles \cite{Kamikado:2013pya}.

\subsection{Correlators}

The most general decomposition of a correlator  is given by
\begin{equation}
\Pi=\Pi_S + i\gamma_5 \Pi_P + \gamma_\mu\Pi_V^\mu + \gamma_\mu \gamma_5 \Pi_A^\mu + \sigma_{\mu\nu}\Pi_T^{\mu\nu} \,,
\label{eq_Clifford}
\end{equation}
where the sub-indices refer, respectively, to scalar, pseudoscalar, vector, axial-vector and tensor structures. 
If no topological anomalies are present, then $\Pi_P=0$.
In the vacuum, there are only scalar and vector contributions, with $\Pi_S=\Pi_2$ and $\Pi_V^\mu=p^\mu\Pi_1$. 
In the presence of an external uniform-electromagnetic field, the most general combinations of the vector, tensor and axial-vector components involve $p_\mu$, $g_{\mu\nu}$, $\epsilon_{\mu\nu\alpha\beta}$ and $F_{\mu\nu}$, on account of the correlator being parity even. 
For the case of a constant  external magnetic field aligned along the third coordinate, the electromagnetic strength tensor can be written as $F_{\mu\nu}=B\epsilon^\perp_{\mu\nu}$, where the perpendicular, anti-symmetric tensor is defined as $\epsilon^\perp_{\mu\nu}\equiv\epsilon_{0\mu\nu 3}$.
The most general decompositions of the vector, axial-vector and tensor structures in Eq. (\ref{eq_Clifford}) are given by
\begin{align}
    \Pi_V^\mu  &=  p_\parallel^\mu\,\Pi_V^\parallel+p^\mu_\perp\ \Pi_V^\perp+\tilde p_\perp^\mu \tilde\Pi_V^\perp,
        \label{eq_Piv}\\
    \Pi_A^\mu  &=  \tilde p_\parallel^{\mu}  \Pi_A,
        \label{eq_PiA}\\
\Pi_T^{\mu\nu} & = \epsilon_\perp^{\mu\nu}\Pi_T^\perp 
    + (p_\parallel^\mu p_\perp^\nu - p_\parallel^\nu p_\perp^\mu)\Pi_T^{\parallel\perp}
    \nonumber \\ & \hspace{2cm} 
    + (p_\parallel^\mu \tilde p_\perp^\nu -  p_\parallel^\nu \tilde p_\perp^\mu)\tilde\Pi_T^{\parallel\perp},
        \label{eq_PiT}
\end{align}
where $\tilde p_\perp^\mu\equiv \epsilon_\perp^{\mu\alpha}p_\alpha$,  $\tilde p_\parallel^\mu\equiv \epsilon_\parallel^{\mu\alpha}p_\alpha$, with the parallel anti-symmetric tensor defined as $\epsilon_\parallel^{\mu\nu} \equiv \frac{1}{2}\epsilon^{\mu\nu\alpha\beta}\epsilon^\perp_{\alpha\beta} = \epsilon^{\mu 12 \nu}$. 
In this work we consider the structures $\Pi_S$, $\Pi_V^\parallel$, $\Pi_V^\perp$ and $\Pi_T^\perp$, in order to compute the FESR.

Usually in the literature the structure decomposition of the correlator is expressed in terms of combinations of external momenta and Dirac matrices \cite{Ioffe:1983ju,Wang:2008vg}, which are related to the above structures by
\begin{align}
F_{\mu\nu}(\slashed{p}\,\sigma^{\mu\nu}+\sigma^{\mu\nu}\slashed{p})\label{eq_Pi_munu-1}
&= 4B \gamma_\mu\gamma_5 \,\tilde p_\parallel^\mu,\\
iF_{\mu\nu}(p^\mu \gamma^\nu-p^\nu \gamma^\mu)\slashed{p} &= 
B\sigma_{\mu\nu}(\tilde p^\mu_\perp p^\nu_\parallel - \tilde p^\nu_\perp p^\mu_\parallel + \epsilon^{\mu\nu}_\perp p_\perp^2),\label{eq_Pi_munu-2}
\end{align}
which is the axial-vector function  $\Pi_A$ in Eq.\,(\ref{eq_Pi_munu-1}), and  a combination of tensor components $\tilde\Pi_T^{\parallel\perp}$ and $\Pi_T^\perp$ in Eq.\,(\ref{eq_Pi_munu-2}).

\subsection{QCD contour integrals}    

The advantage of FESR lies on the natural truncation of the non-perturbative OPE series when integrating around the circle in Eq. (\ref{FESR}).
To visualize this truncation in the presence of external magnetic fields we denote by   $\Pi_n$ the  correlator term of  order $B^n$. 
Then, one has $\Pi_n(s) \sim \frac{\partial^n}{\partial s^n}\Pi_0(s)$. 
This is easily understood in the chiral limit \cite{Ayala:2015qwa} as the only scales available are $s$ and $eB$, and the corrections to the vacuum correlator will be of order $(eB/s)^n$. 
However, there are infrared divergences, which can be handled by including quarks masses. 
In fact, quarks acquire magnetic masses even in the chiral limit \cite{Dominguez:2018njv,Villavicencio:2020fcz}. 

Considering non-logarithmic contributions, the general structure of a Feynman diagram will be $\Pi(s)\sim\int_{x_i}f(x_i)/[s-{\cal M}^2]^n$, where $f$ is some analytic function, $n$ is any integer number, and the integral in terms of Feynman parameters $x_i$ is defined as $\int_{x_i} = \int_0^1 dx_1\dots dx_k\delta(x_1+\dots+x_k-1)$.
The mass term, if one considers equal quark masses, is defined as ${\cal M}^2\equiv m_q^2\left(\frac{1}{x_1}+\dots+\frac{1}{x_k}\right)$.
Integrating around the circle of radius $s_0$ before integrating in the Feynman parameters,  one obtains 
\begin{multline}
\oint_{s_0} \frac{ds}{2\pi i}  s^{N-1}
\left[\int_{x_i}\frac{f(x_i)}{[s - {\cal M}^2]^n}\right]=\\
\theta_{N-n}\binom{N-1}{n-1}\int_{x_i}\theta(s_0-{\cal M}^2) f(x_i)({\cal M}^2)^{N-n}
, \label{eq_OInt_s}
\end{multline}
where the usual step-function is denoted as $\theta(\xi)$, and we define a discrete theta function $\theta_{j}=1$ for $j\geq 0$ and $\theta_{j}=0$ for $j< 0$.
Therefore, the magnetic field powers that participate are limited by the  powers of $s$ in the FESR kernel.

The other kind of integrals contain logarithms. Hence, we first  integrate around the contour obtaining
\begin{multline}
\oint_{s_0} \frac{ds}{2\pi i}  s^{N-1}\left[\int_{x_i}f(x_i) s^n \ln(-s + {\cal M}^2)\right]=\\
\frac{1}{N+n}\int_{x_i}\theta(s_0-{\cal M}^2) f(x_i)[s_0^{N+n}-({\cal M}^2)^{N+n}]
, \label{eq_OInt_ln}
\end{multline}
valid of course for $N+n>0$.

This is a very useful technique to handle infrared divergences without the need for integrating the Feynman parameters in the full correlator. 
Once we integrated along the contour, the Heaviside function $\theta(s_0-{\cal M}^2)$ will provide limits of integration to the Feynman parameters.
Hence, an expansion in terms of $m_q^2/s_0$ can be performed to the lowest order leading to logarithmic corrections.
These logarithmic contributions are strictly magnetic, thus vanishing for $B=0$. 
More details are given in Appendix\,\ref{app_Fparam}.

From the above magnetic power counting, and for sum rules with  kernel $K=1$,  it follows that the perturbative part in Eq.\,(\ref{eq_FESR-Pi1}), and the dimension $d= 3$ condensate contribution  in Eq.\,(\ref{eq_FESR-Pi2}), will have direct magnetic contributions. 

It is important to mention that although there is no contribution of dimension-five  operators in vacuum, there will be such contributions in a magnetic field. For instance,  this happens in the analysis of  \cite{Ioffe:1983ju} involving the axial-vector and the tensor structures in  Eq.\,(\ref{eq_Clifford}). However, they are expected to be negligible for the the magnetic field strengths consider here.

\section{Results}
\label{sec_results}

The magnetic FESR  involving  $\Pi_V^\parallel$, $\Pi_V^\perp$, $\Pi_S$ and $\Pi_T^\perp$, in  the frame $p_\perp =0$, are as follows:
\begin{align}
\lambda_p^2 &=  \frac{s_p^3}{192\pi^4}  
    + \frac{s_p}{32\pi^3}\langle\alpha_s G^2\rangle 
    + \frac{2}{3}\langle\bar uu\rangle^2 
%    \nonumber\\&\quad
    +\frac{s_p}{2\pi^4}e_u e_d B^2
    \nonumber\\&\quad
    +\frac{ s_p}{6\pi^4}(e_uB)^2[\ln(s_p/8m_q^2)-1]
    \nonumber\\&\quad
    +\frac{s_p}{96\pi^4}(e_dB)^2\left[8\ln(s_p/8m_q^2)-9\right],\label{eq_FESRB-1}
%    \\&\nonumber\\
\end{align}
\begin{align}
\lambda_p^2 v_p &=  \frac{s_p^3}{192\pi^4}
    + \frac{s_p}{32\pi^3}\langle\alpha_s G^2\rangle 
    + \frac{2}{3}\langle\bar uu\rangle^2 
    +\frac{s_p}{4\pi^4}e_u e_d B^2
    \nonumber\\&\quad
    -\frac{ s_p}{6\pi^4}(e_uB)^2[\ln(s_p/8m_q^2)-1]
    \nonumber\\&\quad
    -\frac{s_p}{96\pi^4}(e_dB)^2\left[8\ln(s_p/8m_q^2)-9\right],
        \label{eq_FESRB-2}
%    \\&\nonumber\\
\end{align}
\begin{align}
\lambda_p^2 m_p  &= -\frac{s_p^2}{8\pi^2}\langle\bar dd\rangle
    +\frac{1}{12\pi}\langle \alpha_sG^2\rangle\langle\bar dd\rangle
    +\frac{s_p}{2\pi^2}e_uB\langle \bar d\sigma_{12}d\rangle
    \nonumber\\&\quad
    +\frac{4}{3\pi^2}(e_u B)^2 \left[\ln(s_p/m_q^2)-1\right]\langle\bar dd\rangle,
        \label{eq_FESRB-3}
%    \\&\nonumber\\
\end{align}
\begin{align}
-\lambda_p^2 \frac{\kappa_p B}{2} &= \frac{s_p^2}{48\pi^2}\langle \bar d\sigma_{12}d\rangle
   + \frac{e_u B\,s_p}{24\pi^2}\langle\bar dd\rangle,\hspace{1.6cm}
   \label{eq_FESRB-4}
\end{align}
respectively, where $s_p$ is the proton continuum threshold, $\lambda_p$ is the current-proton coupling, $v_p$ is the proton transverse velocity and $\kappa_p$ the proton anomalous magnetic moment.

The FESR for the neutron correlator are the same as above, except for the  change in flavor $d\leftrightarrow u$ and $p\leftrightarrow n$.
Details of the derivation of the above results are given in Appendix\,\ref{app_correlators}.\\

The only new entirely magnetic condensate is the spin polarization one $\langle\bar q\sigma_{12}q\rangle$, often referred to as the anomalous magnetic moment condensate.
From Eq.\, (\ref{eq_FESRB-4}) it is easy to obtain the quark susceptibility at zero magnetic field, defined as $\chi_q=\langle \bar q\sigma_{12}q\rangle/e_qB\langle\bar qq\rangle$.
At $B=0$, Eq.\,(\ref{eq_FESRB-4}) gives $\chi_d(0) = -5.50$\,GeV$^{-2}$ from proton FESR, while using neutron FESR we find $\chi_d(0) = -3.83$\,GeV$^{-2}$.
These results are in a good agreement with expectations  (see \cite{Frasca:2011zn} for a literature review on  $\chi_q$). The main difference with  other determinations is that we do not assume flavor independence. Averaging these two results gives $\bar\chi_q(0)=-4.67$\,GeV$^{-2}$.

\subsection{Inputs}

The first input is the quark condensate as a function of an external magnetic field, as obtained in \cite{Bali:2012zg}, fitted with a Pad\'e approximant. Next, the gluon condensate determined in \cite{Dominguez:2018njv, DElia:2015eey} shows a minor dependence on the magnetic field, so it is assumed constant. Next, 
the (average) light quark mass in vacuum, following from the Gell-Mann-Oakees-Renner relation, is  $m_q(0)=6.05$ MeV, at a scale of $1 \mbox{GeV}$. The running values from logarithmic terms have negligible importance in the presence of magnetic fields. 
Finally, for the third input we consider two scenarios. 
The first one is to take as  input the nucleon mass in order to obtain the condensates $\langle \bar q\sigma_{12} q\rangle$. The second possibility is to consider a constant magnetic susceptibility of the quark condensate in order to obtain the nucleon mass.

For the nucleon mass at finite magnetic field there are different behaviors in the literature. 
Effective models such as the Walecka model, or the linear sigma model or other quark-hadron models, give either increasing masses with increasing magnetic field \cite{Haber:2014ula,Mukherjee:2018ebw}, or no significant variations \cite{Yue:2008tp}. Other models, assuming hadron masses as the sum of constituent quark  masses, lead to a decreasing nucleon mass \cite{Taya:2014nha,Endrodi:2019whh}. Given that all these different evolutions do no change drastically the final results we keep the nucleon mass constant.

In the case  of a constant magnetic susceptibility, one can obtain the magnetic dependence of the nucleon mass for weak magnetic fields. 

\subsection{Numerical results}

We proceed to solve the set of four equations obtained from the FESR,  Eqs.\,(\ref{eq_FESRB-1})-(\ref{eq_FESRB-4}), considering first a constant nucleon mass.

\begin{figure}%[b]
\includegraphics[scale=.67]{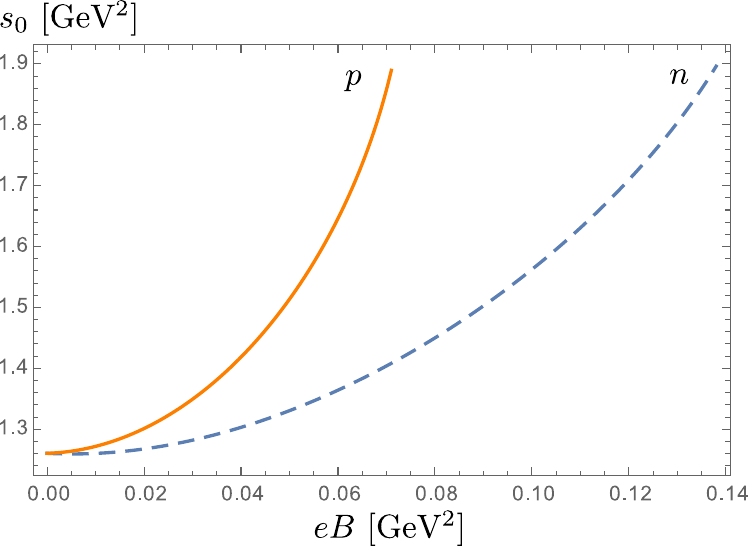}
\caption{Continuum hadronic thresholds $s_0$ as a function of the magnetic field for the proton (solid line) and the neutron correlator (dashed line).}
\label{fig_s0}
\end{figure}

Figure\,\ref{fig_s0} shows the continuum hadronic threshold, $s_0(eB)$ for the nucleon correlators, i.e. for proton and neutron. 
The upper value of $s_0$ is chosen below the production threshold of the nucleon resonance $N^*(1440)$, which is not part of this analysis. This truncation is a standard, and convenient procedure in QCD FESR applications.
Since the pole mass of the  $N^*(1440)$ is about 1.37\,GeV \cite{Tanabashi:2018oca}, then we consider $eB < 0.07$\,GeV$^2$ for the proton correlator, and $eB < 0.14$\,GeV$^2$ for the nucleon case.

\begin{figure}%[t]
\includegraphics[scale=.67]{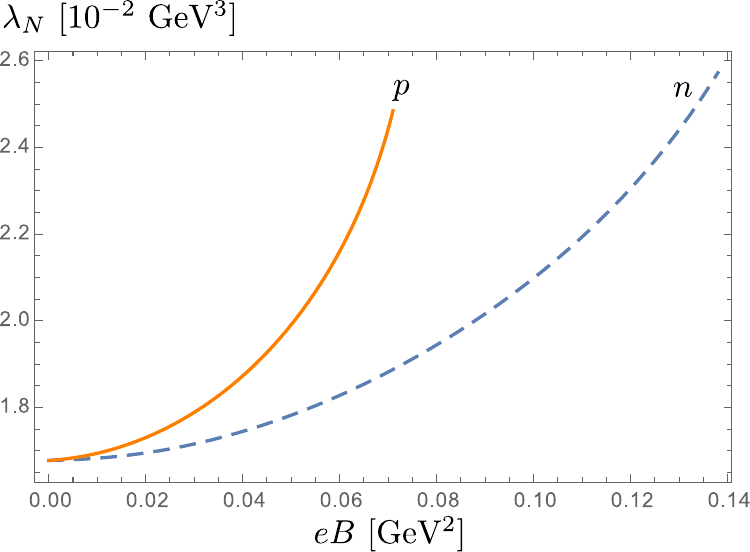}
\caption{Current-nucleon coupling  for the proton (solid line) and the neutron (dashed line) as a function of the magnetic field strength.}
\label{fig_lambda}
\end{figure}

The magnetic behavior of the current-nucleon coupling $\lambda_N$, is shown in Fig.~\ref{fig_lambda} for the proton and neutron. 
We can see that the increasement in the current-nucleon couplings is definitely non negligible, and the change can be larger than 50\% of their initial values. 
The correlator of two nucleonic currents at finite temperature has been explored in \cite{Dominguez:1992vg}.
Contrary to what happens in magnetic fields, the temperature dependence of the  current-nucleon coupling $\lambda_N (T)$ turns out to be a monotonically decreasing function, vanishing at $T_{c}$.

\begin{figure}%[t]
\includegraphics[scale=.67]{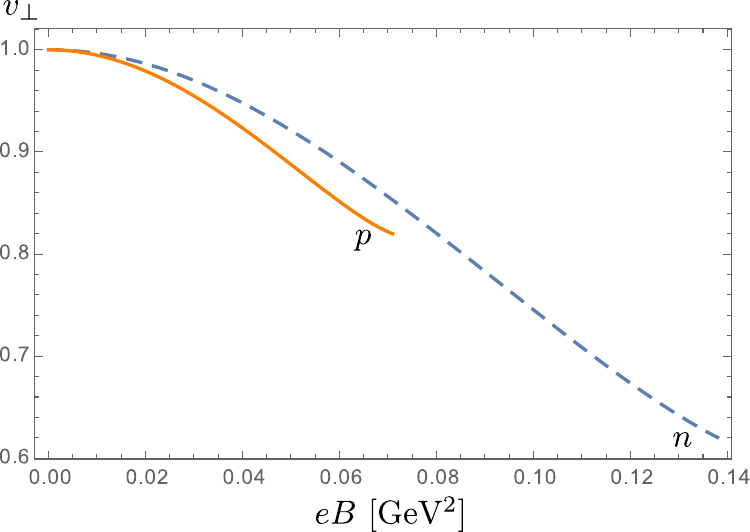}
\caption{Transverse velocity as a function of the magnetic field strength. 
Solid line for proton and dashed line for neutron.}
\label{fig_v}
\end{figure}

In Fig.~\ref{fig_v}, we observe the change of the transverse velocity $v_\perp$ for protons and neutrons, as a function of the magnetic field. 
Both cases have a decreasing behavior and they are always less than the speed of light  as it is expected.

\begin{figure}%[b]
\includegraphics[scale=.67]{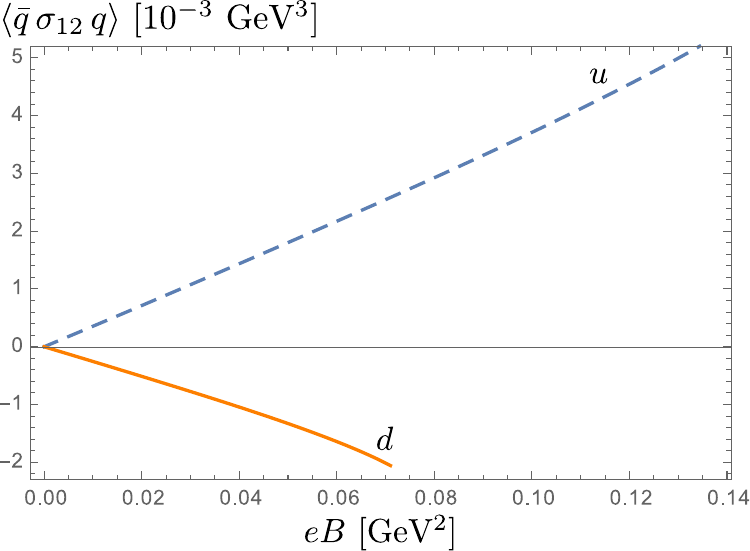}
\caption{Magnetic behavior of the spin polarization condensate (also known as anomalous magnetic moment condensate) $\langle \bar{q}\sigma_{1,2}q \rangle$ as a function of the magnetic field. Shown are up-quark (dashed line) and down-quark (solid line) terms.}
\label{fig_qsq}%[b]
\end{figure}

In the QCD sector, we determinate the magnetic behavior of the spin polarization condensate $\langle \bar q\sigma_{12}q\rangle$, and it is shown in Fig.~\ref{fig_qsq}. The up and down quarks exhibit opposite behavior. 
This is to be expected since they have a different electric charge. 
The behavior of our result is in agreement with  results found in the literature \cite{Frasca:2011zn,Bali:2020bcn}.

\begin{figure}
\includegraphics[scale=.66]{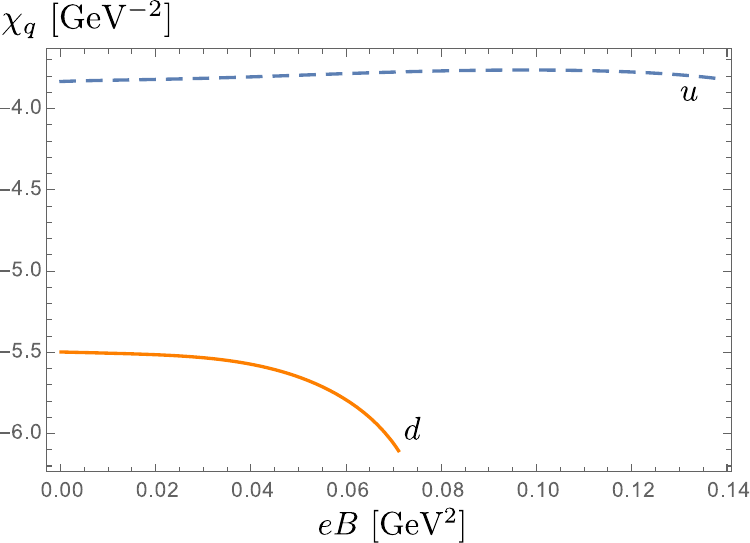}
\caption{Magnetic susceptibility of the quark condensate for the down-quark (solid line) and the up-quark (dashed line) as a function of the magnetic field strength.}
\label{fig_chi}
\end{figure}

The magnetic susceptibility for $u$ and $d$ quarks is essentially constant as can be appreciated in Fig.\,\ref{fig_chi} for almost the whole range of values of the magnetic field strength. 
This means that one can consider a constant magnetic susceptibility as a good approach for low magnetic field. 
If we consider a constant magnetic field dependent nucleon mass, the behavior of the magnetic susceptibility does not have a significant change, remaining almost constant for lower values of the magnetic field. 
This leads us to speculate about what kind of information can be obtained considering a constant magnetic susceptibility, because in this case we have an extra input.

\begin{figure}%[t]
\includegraphics[scale=.66]{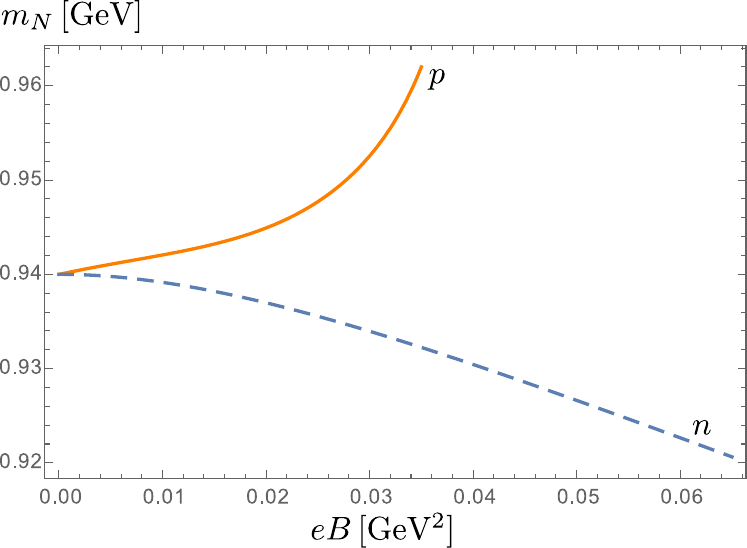}
\caption{Proton and neutron masses as a function of the magnetic field strength.}
\label{fig_mN}
\end{figure}

If we now consider the condensates $\langle \bar q\sigma_{12}q\rangle \approx e_qB \chi_q(0)\langle\bar q q\rangle$ for a given value of the magnetic susceptibility, we can reduce the number of inputs, and therefore obtain  the evolution of nucleon masses at low $B$.
Using the values obtained at $B=0$  from Fig.\,\ref{fig_chi}, i.e. $\chi_d(0) = -5.50$\,GeV$^{-2}$ and $\chi_u(0) = -3.83$\,GeV$^{-2}$,   the resulting nucleon masses are shown in Fig.\,\ref{fig_mN}.
The maximum values of the magnetic field are  obtained by imposing  the variation in  magnetic susceptibility in Fig.\,\ref{fig_chi} to be less than 0.05\,GeV$^2$. 

The behavior of the nucleon masses agrees with that obtained from lattice QCD results \cite{Endrodi:2019whh} leading  to an increasing proton mass and a decreasing neutron mass at low $B$.
However, these results are focused in the high magnetic field region.
The increase of the proton mass agrees with  results obtained in \cite{Mukherjee:2018ebw} from considering the anomalous nucleon magnetic moment.
It is important to remark that the definition of nucleon mass might change substantively in different approaches. 
For that reason we do not compare with results that take the mass as the minimal energy in the lowest Landau level.

It is worth mentioning that for the proton case it is not possible to extend the numerical evaluation for $eB> 0.04$\,GeV$^2$,  thus highlighting the importance of magnetic effects for the full $\langle \bar q\sigma_{12}q\rangle$ condensate.
On the other hand, for $eB>0.12$\,GeV$^2$ the neutron mass exhibits an inflection in the curve and starts to grow with the magnetic field. However at this point, the results are unreliable.

The magnetic behavior of the  nucleon mass is quite different from that at  finite temperature  \cite{Dominguez:1992vg}  where it remains approximately constant in a wide range of temperature,  increasing sharply near the critical deconfinement temperature $T_{c}$.

\section{Conclusions}
\label{sec_conclusions}

Different nucleon parameters were determined using FESR for nucleon-nucleon correlators in the presence of an external uniform magnetic field.
The evolution of the  quark condensates, as function of the magnetic field and the nucleon masses, was considered as an input. 
The magnetic evolution of the quark condensates is one of the main and direct contributions for the magnetic evolution of the other parameters, in contrast to the case of magnetic evolution of nucleon masses which produces a minimal impact.

The obtained threshold which delimits the quark-hadron duality was calculated for protons and neutrons obtaining different growing values.
The maximum values must be smaller than the mass pole of next nucleon resonances in order to avoid the incorporation of them in the spectral function. 
This limit provides a maximum magnetic field allowed in our approach of $eB<0.07$\,GeV$^2\approx 3.6 m_\pi^2$ for the proton channel  and $eB<0.14$\,GeV$^2\approx 7 m_\pi^2$ for the neutron channel.

The quantities that can be obtained are the current-nucleon couplings $\lambda_N$,  which grow with the magnetic field, signaling a stronger confinement of quarks inside nucleons and can increase by more that 50\%.
The transverse velocity $v_\perp$ decreases with the magnetic field as expected, being smaller than the speed of light. 
The magnetic moment condensates $\langle \bar q\sigma_{12} q\rangle$ is another quantity obtained, being in good agreement with the expected behavior from other models. 
From this quantity, the magnetic susceptibility of the quark condensate $\chi_q$ can be obtained, with the interesting behavior that it is almost constant for most of the allowed range of values of the magnetic field. 
The main difference with many other determinations is that we have not considered the same magnetic susceptibility  for the two light flavors, and in fact the result show they are different.

In particular, the magnetic susceptibility at $B=0$ can be compared with many other works and it is in good agreement with the range of values of other results. 
Since $\chi_q(0)$ can be calculated independently of the other sum rules equations, it is possible to obtain the nucleon masses for low magnetic field by keeping $\chi_q(0)$ as an input, valid for low values of the magnetic field.
As a result we obtain a behavior that agrees with lattice results, namely that proton mass increases for low $B$ and neutron mass decreases. 
The increasing proton mass numerically coincides with an analysis in the frame of Walecka model when nucleon anomalous magnetic moment is included.

The increasing of the hadronic threshold and current-nucleon couplings was expected from previous determinations, although every channel in principle behaves independently, but it is an insight of stronger confinement. 
The different behavior of the nucleon masses represents an interesting scenario. 
Although it is valid for a lower magnetic field, we are talking about $eB < 1.8 m_\pi^2$ for protons and $eB < 3.6 m_\pi^2$, it is strong enough for magnetic field in magnetars. 
This is an interesting scenario, worthwhile to be explored including baryonic density effects.

\section*{Acknowledgements}
The authors acknowledge support from FONDECYT (Chile) under grants 1190192, 1170107 and 1200483.
M.L. acknowledges support from ANID/PIA/Basal (Chile) under grant FB0821. L. A. H. acknowledges support from a PAPIIT-DGAPA-UNAM fellowship. C.A.D. was supported in part by the Alexander von Humboldt Foundation (Germany), and the University of Cape Town.
C. V. and M. L. would like to thank N. Scoccola for valuable comments regarding the comparison with lattice results.

\bigskip
\noindent
\emph{Note added}---In [9] the mass reported corresponds to the minimum energy in the lowest Landau level, and the behavior of the proton mass is not clear in the
low magnetic field region we are considering. 
If we take the minimum energy, the tendency then is to diminish with the magnetic field for both nucleons, in agreement with lattice results.

\appendix

\section{Magnetic correlators}
\label{app_correlators}

\begin{figure}
\includegraphics[scale=1.4]{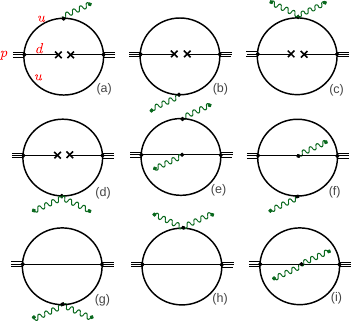}
\caption{Non-vanishing Feynman diagrams with magnetic insertions used in this work.
The diagrams correspond to proton-proton propagator. 
The case of neutron correlator is the same but interchanging $u\leftrightarrow d$ lines.}
\label{fig_diag}
\end{figure}

This appendix deals with current correlator terms in the presence of an external magnetic field.
\emph{Magnetic field insertion} is the name given to the individual propagators from Eq.\,(\ref{eq_SB}). They are expressed in powers of $B/(p^2-m^2)$, and diagrammatically correspond to a single, double, or higher number of external lines, depending on the power of $B$. Figure\,\ref{fig_diag} shows all  diagrams  with magnetic field insertions, including propagators.
The relevant diagrams are the one-loop ones,  corresponding to the dimension-3 contribution in the OPE  (quark condensate),  i.e. diagrams (a) to (d).The perturbative ones in the OPE  are two-loop diagrams  (e) to (i). Notice that we do not include two-loop diagrams with one magnetic field insertion. This is  because they only contribute   to the axial component of the structure decomposition in Eq.\,(\ref{eq_Clifford}). This is not considered here.
Also notice the vanishing of one-loop and two-loop diagrams with one magnetic insertion, with the top $u$-propagator and one magnetic insertion  from the bottom $u$-propagator.
Finally,  we only consider  contributions to the correlators $\Pi_S$, $\Pi_V$ and $\Pi_T$ from Eq.\,(\ref{eq_Clifford}), as  well as $\Pi_V^\parallel$ and $\Pi_V^\perp$ from Eq.\,(\ref{eq_Piv}), and  $\Pi_T^\perp$ from Eq.\,(\ref{eq_PiT}). Results are expressed in the frame $p_\perp^2=0$, with $p_\parallel^2=s$.

\subsection{One-loop diagrams}

The  correlator corresponding to  diagrams involving operators of dimension $d=3$ is given by
\begin{multline}
    \Pi = i \int\frac{d^4k}{(2\pi)^4} \text{tr}[\gamma_\mu S_u(k) \gamma_\nu C S_u(q-k) C]\\
    \times\gamma^\mu [\langle \bar{d} d \rangle + \sigma_{12} \langle \bar{d} \sigma_{12} d \rangle ] \gamma^\nu.
\end{multline}
This correlator corresponds to the proton  diagrams  (a) to (d) in Fig\,\ref{fig_diag}. 
Notice that the only condensate is the one in the $d$-line. 
This is because when cutting the top or bottom $u$-lines the correlator vanishes after performing the trace. 

The first new contribution is that of the spin polarization condensate, also entering the diagram without magnetic field insertion. 
It contributes to the tensor part in the structure decomposition in Eq.\,(\ref{eq_PiT}),
\begin{equation}
    \Pi_{T(0)}^{\perp } = -\frac{1}{24\pi^2}s\ln (-s/\nu^2)\langle\bar d\, \sigma_{12}\,d\rangle,
\end{equation}
where the subscript $(0)$ indicates that it is of order $B^0$.
This contribution  does not involve magnetic insertions, with magnetic effects implicit in the spin polarization condensate. 

Diagrams (a)+(b) give a contribution to a scalar and to a tensor component
\begin{align}
\Pi_{S(a+b)} &=- \frac{e_u B}{2\pi^2} \ln(-s/\nu^2)\langle \bar{d}\sigma_{12}d \rangle, \\
 \Pi_{T(a+b)}^\perp &=\frac{e_u B}{4\pi^2} \ln(-s/\nu^2)\langle \bar{d}d \rangle.
\end{align}

Diagrams (c)+(d) generate a contribution to a tensor component, not considered here,  and a contribution to the scalar component which gives
\begin{equation}
  \Pi_{S(c+d)}=-\frac{4}{\pi^2}(e_u B)^2 \int_x \frac{1-x}{3x} \frac{1}{p^2-{\cal M}_1^2} \langle\bar dd\rangle,
  \label{eq_Pi_S_cd}
\end{equation}
with ${\cal M}_1=m_q^2/x(1-x)$. This last equation vanishes in the chiral limit. Hence, it is not necessary to keep the quark mass in the denominator to regularize it.
Details of the integration in  Feynman parameters is given in Appendix\,\ref{app_Fparam}.

\subsection{Two-loop diagrams}

The  perturbative correlator corresponding to  two-loop diagrams  is 
\begin{multline}
    \Pi=12 i \int \frac{d^4k}{(2\pi)^4} \frac{d^4p}{(2\pi)^4}
    \text{tr}[\gamma_\mu S_u(k) \gamma_\nu C S_u(p) C] \\
    \times \gamma^\mu S_d(q-k-p) \gamma^\nu.
\end{multline}
Diagrams (e) and (f), as well as diagrams (g) and (h) give the same result. All  diagrams from (e) to (i)  contribute only to the vector component of the correlator.

The parallel vector components are
\begin{align}
    \Pi_{V(e+f)}^\parallel &= -\frac{e_u e_d B^2}{2\pi^4}\ln(-s),\\
       &\nonumber\\
   \Pi_{V(g+h)}^\parallel &= -\frac{(e_uB)^2}{\pi^4}\int_0^1 dx dy \ln(-s+{\cal M}_2^2)
   \nonumber\\&\hspace{3cm}
   \times \frac{x y^4(1-y) z}{(xy+yz+zx)^5},\label{eq_Pi_V_par-gh}\\
        &\nonumber\\
    \Pi_{V(i)}^\parallel &= -\frac{(e_dB)^2}{4\pi^4}\int_0^1 dx dy \ln(-s+{\cal M}_2^2)
    \nonumber\\&\hspace{2cm}
    \times  \frac{2xy^4(1-y)z-x^2y^3z^2}{(xy+yz+zw)^5},\label{eq_Pi_V_par-i}
\end{align}
with ${\cal M}_2^2=m_q^2(\frac{1}{x}+\frac{1}{y}+\frac{1}{z})$ and $z=1-x-y$. 

The perpendicular vector components are
\begin{align}
    \Pi_{V(e+f)}^\perp &=\frac{1}{2} \Pi_{V(e+f)}^\parallel,\\
    \Pi_{V(g+h)}^\perp &= -\Pi_{V(g+h)}^\parallel,\\
    \Pi_{V(i)}^\perp &= -\Pi_{V(i)}^\parallel.
\end{align}

The quark mass entering the logarithms of diagrams (g), (h), and (i) is kept finite in order to regularize  infrared divergences. 

\section{Feynman parameters}
\label{app_Fparam}

Integrating the QCD contributions on the contour in the $s$-plane leads to integrals involving Feynman parameters, as  described in Eqs. (\ref{eq_OInt_s}) and (\ref{eq_OInt_ln}).
The quark mass regularizes infrared divergences entering diagrams  Fig.\,\ref{fig_diag} (c), (d), (g) and (h), as described next.
  
\subsection{One-loop diagrams}

Diagrams with a quark condensate are one-loop. 
From Eq.\, (\ref{eq_OInt_s})  they are written as
\begin{equation}
    \oint_{s_0} \frac{ds}{2\pi i}\Pi_{\text{1-loop}}=\int_0^1 dx dy\,\delta(x+y-1)\theta(s_0-{\cal M}_1^2)f(x,y),
\end{equation}
with ${\cal M}_1^2 = m_q^2(\frac{1}{x}+\frac{1}{y})$,  and $f(x,y)$ an analytic function. 
Integrating in $y$, with  the restrictions from the $\delta$- and the $\theta$-function, gives
\begin{equation}
    \oint_{s_0} \frac{ds}{2\pi i}\Pi_{\text{1-loop}}=\theta(s_0-4m_q^2)\int_{x_-}^{x_+} dx f(x,1-x),
\end{equation}
where the integration limits for equal quark masses are $x_\pm=\frac{1}{2}\left[1\pm\sqrt{1-4m_q^2/s_0}\right]$. 
For $m_q\to 0$ one recovers the usual limits $x_-=0$ and $x_+=1$.

The specific one-loop diagrams needed for this regularization are (c) and (d) in Fig.\,\ref{fig_diag}. 
Integrating Eq.\,(\ref{eq_Pi_S_cd}) on the contour in $s$, and then expanding in $m_q/s_0\to 0$ up to first order  gives
\begin{equation}
        \oint_{s_0} \frac{ds}{2\pi i}\Pi_{S(c+d)} = -\frac{4}{3\pi^2}(e_uB)^2\left[\ln(s_p/m_q^2)-1\right]\langle \bar d d\rangle.
\end{equation}

\subsection{Two-loop diagrams}

The perturbative diagrams at  the two-loop level involve logarithmic terms and can be written as
\begin{multline}
    \oint_{s_0} \frac{ds}{2\pi i}\Pi_{\text{2-loop}} =
    \int_0^1 dx dydz\,\delta(x+y+z-1)\\
    \times\theta(s_0-{\cal M}_2^2)f(x,y,z)[s_0-{\cal M}_2^2],
\end{multline}
with ${\cal M}_2^2 = m_q^2(\frac{1}{x}+\frac{1}{y}+\frac{1}{z})$ and $f(x,y,z)$ an analytic function.
Integrating in $z$, and considering the restrictions from the $\delta$- and  $\theta$-functions, gives
\begin{multline}
    \oint_{s_0} \frac{ds}{2\pi i}\Pi_{\text{2-loop}}=\theta(s_0-9m_q^2)\\
    \times\int_{x_-}^{x_+} \! dx \int_{y_-}^{y_+}\! dy \;f(x,y,z) [s_0-{\cal M}_2^2],
\end{multline}
with $z=1-x-y$ and
\begin{align}
    x_\pm &= \frac{1}{2}\left[ 1- \frac{3m^2}{s_0}\pm
    \sqrt{\left(1-\frac{3m^2}{s_0}\right)^2-\frac{4m^2}{s_0}}\right],\\
    y_\pm &= \frac{1}{2}\left[ 1- x \pm
    \sqrt{(1-x)^2-\frac{4m^2\,x(1-x)}{s_0 \,x-m^2}}\right].
\end{align}
In the limit $m_q\to 0$  one recovers the usual limits $x_-=0$, $x_+=1$, $y_-=0$ and $y_+=1-x$.
There are four non-trivial integrals involved here, as can be seen from Eqs. (\ref{eq_Pi_V_par-gh}) and  (\ref{eq_Pi_V_par-i}).
All these integrals can be evaluated analytically when integrating in the variable  $y$, but not in $x$ which will require some fitting.
Starting with the last term in Eq.\,(\ref{eq_Pi_V_par-i}), in the limit $m_q\to 0$, which becomes
\begin{multline}
   \int_{x_-}^{x_+}dx\int_{y_-}^{y_+} \frac{x^2 y^3 z^2}{(xy+yz+zx)^5} [s_p-{\cal M}_2^2]  = \frac{s_p}{12}.
\end{multline}
The term in this integral proportional to $s_p$ can be obtained directly in the chiral limit. 
The term proportional to ${\cal M}_2^2$ is evaluated numerically, but it vanishes in the limit $m_q\to 0$. 
Next term is the one that enters in both equations (\ref{eq_Pi_V_par-gh}) and  (\ref{eq_Pi_V_par-i}). 
In the limit $m_q\to 0$ it becomes
\begin{multline}
   \int_{x_-}^{x_+}dx\int_{y_-}^{y_+} \frac{x y^4(1-y) z}{(xy+yz+zx)^5} [s_p-{\cal M}_2^2]\\  \approx \frac{s_p}{6}\left[\ln(s_p/8m_q^2)-1\right],
\end{multline}
where it  corresponds to a fit of the result expressed as an integral in $x$,  as a function of $m_q^2/s_p$.
From these considerations, the contour integral of Eqs. (\ref{eq_Pi_V_par-gh}) and  (\ref{eq_Pi_V_par-i}) leads to
\begin{align}
     \oint_{s_p} \frac{ds}{2\pi i}\Pi^\parallel_{V(g+h)} &= -\frac{(e_uB)^2}{12\pi^4}s_p\left[\ln(s_p/8m_q^2)-1\right],\\
     \oint_{s_p} \frac{ds}{2\pi i}\Pi^\parallel_{V(i)} &= -\frac{(e_dB)^2}{96\pi^4}s_p\left[8\ln(s_p/8m_q^2)-9\right].
\end{align}

\bibliographystyle{apsrev4-1}
\bibliography{nucleon}

%merlin.mbs apsrev4-1.bst 2010-07-25 4.21a (PWD, AO, DPC) hacked
%Control: key (0)
%Control: author (72) initials jnrlst
%Control: editor formatted (1) identically to author
%Control: production of article title (-1) disabled
%Control: page (0) single
%Control: year (1) truncated
%Control: production of eprint (0) enabled
\begin{thebibliography}{31}%
\makeatletter
\providecommand \@ifxundefined [1]{%
 \@ifx{#1\undefined}
}%
\providecommand \@ifnum [1]{%
 \ifnum #1\expandafter \@firstoftwo
 \else \expandafter \@secondoftwo
 \fi
}%
\providecommand \@ifx [1]{%
 \ifx #1\expandafter \@firstoftwo
 \else \expandafter \@secondoftwo
 \fi
}%
\providecommand \natexlab [1]{#1}%
\providecommand \enquote  [1]{``#1''}%
\providecommand \bibnamefont  [1]{#1}%
\providecommand \bibfnamefont [1]{#1}%
\providecommand \citenamefont [1]{#1}%
\providecommand \href@noop [0]{\@secondoftwo}%
\providecommand \href [0]{\begingroup \@sanitize@url \@href}%
\providecommand \@href[1]{\@@startlink{#1}\@@href}%
\providecommand \@@href[1]{\endgroup#1\@@endlink}%
\providecommand \@sanitize@url [0]{\catcode `\\12\catcode `\$12\catcode
  `\&12\catcode `\#12\catcode `\^12\catcode `\_12\catcode `\%12\relax}%
\providecommand \@@startlink[1]{}%
\providecommand \@@endlink[0]{}%
\providecommand \url  [0]{\begingroup\@sanitize@url \@url }%
\providecommand \@url [1]{\endgroup\@href {#1}{\urlprefix }}%
\providecommand \urlprefix  [0]{URL }%
\providecommand \Eprint [0]{\href }%
\providecommand \doibase [0]{http://dx.doi.org/}%
\providecommand \selectlanguage [0]{\@gobble}%
\providecommand \bibinfo  [0]{\@secondoftwo}%
\providecommand \bibfield  [0]{\@secondoftwo}%
\providecommand \translation [1]{[#1]}%
\providecommand \BibitemOpen [0]{}%
\providecommand \bibitemStop [0]{}%
\providecommand \bibitemNoStop [0]{.\EOS\space}%
\providecommand \EOS [0]{\spacefactor3000\relax}%
\providecommand \BibitemShut  [1]{\csname bibitem#1\endcsname}%
\let\auto@bib@innerbib\@empty
%</preamble>
\bibitem [{\citenamefont {Kharzeev}\ \emph {et~al.}(2013)\citenamefont
  {Kharzeev}, \citenamefont {Landsteiner}, \citenamefont {Schmitt},\ and\
  \citenamefont {Yee}}]{Kharzeev:2013jha}%
  \BibitemOpen
  \bibinfo {editor} {\bibfnamefont {D.}~\bibnamefont {Kharzeev}}, \bibinfo
  {editor} {\bibfnamefont {K.}~\bibnamefont {Landsteiner}}, \bibinfo {editor}
  {\bibfnamefont {A.}~\bibnamefont {Schmitt}}, \ and\ \bibinfo {editor}
  {\bibfnamefont {H.}~\bibnamefont {Yee}},\ eds.,\ \href {\doibase
  10.1007/978-3-642-37305-3} {\emph {\bibinfo {title} {{Strongly Interacting
  Matter in Magnetic Fields}}}}\ (\bibinfo  {publisher} {Springer},\ \bibinfo
  {year} {2013})\BibitemShut {NoStop}%
\bibitem [{\citenamefont {Dominguez}\ \emph {et~al.}(2018)\citenamefont
  {Dominguez}, \citenamefont {Loewe},\ and\ \citenamefont
  {Villavicencio}}]{Dominguez:2018njv}%
  \BibitemOpen
  \bibfield  {author} {\bibinfo {author} {\bibfnamefont {C.~A.}\ \bibnamefont
  {Dominguez}}, \bibinfo {author} {\bibfnamefont {M.}~\bibnamefont {Loewe}}, \
  and\ \bibinfo {author} {\bibfnamefont {C.}~\bibnamefont {Villavicencio}},\
  }\href {\doibase 10.1103/PhysRevD.98.034015} {\bibfield  {journal} {\bibinfo
  {journal} {Phys. Rev. D}\ }\textbf {\bibinfo {volume} {98}},\ \bibinfo
  {pages} {034015} (\bibinfo {year} {2018})},\ \Eprint
  {http://arxiv.org/abs/1806.10088} {arXiv:1806.10088 [hep-ph]} \BibitemShut
  {NoStop}%
\bibitem [{\citenamefont {D'Elia}\ \emph {et~al.}(2016)\citenamefont {D'Elia},
  \citenamefont {Meggiolaro}, \citenamefont {Mesiti},\ and\ \citenamefont
  {Negro}}]{DElia:2015eey}%
  \BibitemOpen
  \bibfield  {author} {\bibinfo {author} {\bibfnamefont {M.}~\bibnamefont
  {D'Elia}}, \bibinfo {author} {\bibfnamefont {E.}~\bibnamefont {Meggiolaro}},
  \bibinfo {author} {\bibfnamefont {M.}~\bibnamefont {Mesiti}}, \ and\ \bibinfo
  {author} {\bibfnamefont {F.}~\bibnamefont {Negro}},\ }\href {\doibase
  10.1103/PhysRevD.93.054017} {\bibfield  {journal} {\bibinfo  {journal} {Phys.
  Rev. D}\ }\textbf {\bibinfo {volume} {93}},\ \bibinfo {pages} {054017}
  (\bibinfo {year} {2016})},\ \Eprint {http://arxiv.org/abs/1510.07012}
  {arXiv:1510.07012 [hep-lat]} \BibitemShut {NoStop}%
\bibitem [{\citenamefont {Frasca}\ and\ \citenamefont
  {Ruggieri}(2011)}]{Frasca:2011zn}%
  \BibitemOpen
  \bibfield  {author} {\bibinfo {author} {\bibfnamefont {M.}~\bibnamefont
  {Frasca}}\ and\ \bibinfo {author} {\bibfnamefont {M.}~\bibnamefont
  {Ruggieri}},\ }\href {\doibase 10.1103/PhysRevD.83.094024} {\bibfield
  {journal} {\bibinfo  {journal} {Phys. Rev. D}\ }\textbf {\bibinfo {volume}
  {83}},\ \bibinfo {pages} {094024} (\bibinfo {year} {2011})},\ \Eprint
  {http://arxiv.org/abs/1103.1194} {arXiv:1103.1194 [hep-ph]} \BibitemShut
  {NoStop}%
\bibitem [{\citenamefont {Yue}\ and\ \citenamefont {Shen}(2008)}]{Yue:2008tp}%
  \BibitemOpen
  \bibfield  {author} {\bibinfo {author} {\bibfnamefont {P.}~\bibnamefont
  {Yue}}\ and\ \bibinfo {author} {\bibfnamefont {H.}~\bibnamefont {Shen}},\
  }\href {\doibase 10.1103/PhysRevC.77.045804} {\bibfield  {journal} {\bibinfo
  {journal} {Phys. Rev. C}\ }\textbf {\bibinfo {volume} {77}},\ \bibinfo
  {pages} {045804} (\bibinfo {year} {2008})},\ \Eprint
  {http://arxiv.org/abs/0804.3027} {arXiv:0804.3027 [nucl-th]} \BibitemShut
  {NoStop}%
\bibitem [{\citenamefont {Haber}\ \emph {et~al.}(2014)\citenamefont {Haber},
  \citenamefont {Preis},\ and\ \citenamefont {Schmitt}}]{Haber:2014ula}%
  \BibitemOpen
  \bibfield  {author} {\bibinfo {author} {\bibfnamefont {A.}~\bibnamefont
  {Haber}}, \bibinfo {author} {\bibfnamefont {F.}~\bibnamefont {Preis}}, \ and\
  \bibinfo {author} {\bibfnamefont {A.}~\bibnamefont {Schmitt}},\ }\href
  {\doibase 10.1103/PhysRevD.90.125036} {\bibfield  {journal} {\bibinfo
  {journal} {Phys. Rev. D}\ }\textbf {\bibinfo {volume} {90}},\ \bibinfo
  {pages} {125036} (\bibinfo {year} {2014})},\ \Eprint
  {http://arxiv.org/abs/1409.0425} {arXiv:1409.0425 [nucl-th]} \BibitemShut
  {NoStop}%
\bibitem [{\citenamefont {Taya}(2015)}]{Taya:2014nha}%
  \BibitemOpen
  \bibfield  {author} {\bibinfo {author} {\bibfnamefont {H.}~\bibnamefont
  {Taya}},\ }\href {\doibase 10.1103/PhysRevD.92.014038} {\bibfield  {journal}
  {\bibinfo  {journal} {Phys. Rev. D}\ }\textbf {\bibinfo {volume} {92}},\
  \bibinfo {pages} {014038} (\bibinfo {year} {2015})},\ \Eprint
  {http://arxiv.org/abs/1412.6877} {arXiv:1412.6877 [hep-ph]} \BibitemShut
  {NoStop}%
\bibitem [{\citenamefont {Mukherjee}\ \emph {et~al.}(2018)\citenamefont
  {Mukherjee}, \citenamefont {Ghosh}, \citenamefont {Mandal}, \citenamefont
  {Sarkar},\ and\ \citenamefont {Roy}}]{Mukherjee:2018ebw}%
  \BibitemOpen
  \bibfield  {author} {\bibinfo {author} {\bibfnamefont {A.}~\bibnamefont
  {Mukherjee}}, \bibinfo {author} {\bibfnamefont {S.}~\bibnamefont {Ghosh}},
  \bibinfo {author} {\bibfnamefont {M.}~\bibnamefont {Mandal}}, \bibinfo
  {author} {\bibfnamefont {S.}~\bibnamefont {Sarkar}}, \ and\ \bibinfo {author}
  {\bibfnamefont {P.}~\bibnamefont {Roy}},\ }\href {\doibase
  10.1103/PhysRevD.98.056024} {\bibfield  {journal} {\bibinfo  {journal} {Phys.
  Rev. D}\ }\textbf {\bibinfo {volume} {98}},\ \bibinfo {pages} {056024}
  (\bibinfo {year} {2018})},\ \Eprint {http://arxiv.org/abs/1809.07028}
  {arXiv:1809.07028 [hep-ph]} \BibitemShut {NoStop}%
\bibitem [{\citenamefont {Endr{\"o}di}\ and\ \citenamefont
  {Mark{\'o}}(2019)}]{Endrodi:2019whh}%
  \BibitemOpen
  \bibfield  {author} {\bibinfo {author} {\bibfnamefont {G.}~\bibnamefont
  {Endr{\"o}di}}\ and\ \bibinfo {author} {\bibfnamefont {G.}~\bibnamefont
  {Mark{\'o}}},\ }\href {\doibase 10.1007/JHEP08(2019)036} {\bibfield
  {journal} {\bibinfo  {journal} {JHEP}\ }\textbf {\bibinfo {volume} {08}},\
  \bibinfo {pages} {036} (\bibinfo {year} {2019})},\ \Eprint
  {http://arxiv.org/abs/1905.02103} {arXiv:1905.02103 [hep-lat]} \BibitemShut
  {NoStop}%
\bibitem [{\citenamefont {Coppola}\ \emph {et~al.}(2020)\citenamefont
  {Coppola}, \citenamefont {Gomez~Dumm},\ and\ \citenamefont
  {Scoccola}}]{Coppola:2020mon}%
  \BibitemOpen
  \bibfield  {author} {\bibinfo {author} {\bibfnamefont {M.}~\bibnamefont
  {Coppola}}, \bibinfo {author} {\bibfnamefont {D.}~\bibnamefont {Gomez~Dumm}},
  \ and\ \bibinfo {author} {\bibfnamefont {N.}~\bibnamefont {Scoccola}},\
  }\href@noop {} {\  (\bibinfo {year} {2020})},\ \Eprint
  {http://arxiv.org/abs/2009.14105} {arXiv:2009.14105 [hep-ph]} \BibitemShut
  {NoStop}%
\bibitem [{\citenamefont {Ioffe}\ and\ \citenamefont
  {Smilga}(1984)}]{Ioffe:1983ju}%
  \BibitemOpen
  \bibfield  {author} {\bibinfo {author} {\bibfnamefont {B.}~\bibnamefont
  {Ioffe}}\ and\ \bibinfo {author} {\bibfnamefont {A.~V.}\ \bibnamefont
  {Smilga}},\ }\href {\doibase 10.1016/0550-3213(84)90364-X} {\bibfield
  {journal} {\bibinfo  {journal} {Nucl. Phys. B}\ }\textbf {\bibinfo {volume}
  {232}},\ \bibinfo {pages} {109} (\bibinfo {year} {1984})}\BibitemShut
  {NoStop}%
\bibitem [{\citenamefont {Wang}\ and\ \citenamefont {Lee}(2008)}]{Wang:2008vg}%
  \BibitemOpen
  \bibfield  {author} {\bibinfo {author} {\bibfnamefont {L.}~\bibnamefont
  {Wang}}\ and\ \bibinfo {author} {\bibfnamefont {F.~X.}\ \bibnamefont {Lee}},\
  }\href {\doibase 10.1103/PhysRevD.78.013003} {\bibfield  {journal} {\bibinfo
  {journal} {Phys. Rev. D}\ }\textbf {\bibinfo {volume} {78}},\ \bibinfo
  {pages} {013003} (\bibinfo {year} {2008})},\ \Eprint
  {http://arxiv.org/abs/0804.1779} {arXiv:0804.1779 [hep-ph]} \BibitemShut
  {NoStop}%
\bibitem [{\citenamefont {Ayala}\ \emph {et~al.}(2015)\citenamefont {Ayala},
  \citenamefont {Dominguez}, \citenamefont {Hernandez}, \citenamefont {Loewe},
  \citenamefont {Rojas},\ and\ \citenamefont {Villavicencio}}]{Ayala:2015qwa}%
  \BibitemOpen
  \bibfield  {author} {\bibinfo {author} {\bibfnamefont {A.}~\bibnamefont
  {Ayala}}, \bibinfo {author} {\bibfnamefont {C.~A.}\ \bibnamefont
  {Dominguez}}, \bibinfo {author} {\bibfnamefont {L.~A.}\ \bibnamefont
  {Hernandez}}, \bibinfo {author} {\bibfnamefont {M.}~\bibnamefont {Loewe}},
  \bibinfo {author} {\bibfnamefont {J.~C.}\ \bibnamefont {Rojas}}, \ and\
  \bibinfo {author} {\bibfnamefont {C.}~\bibnamefont {Villavicencio}},\ }\href
  {\doibase 10.1103/PhysRevD.92.016006} {\bibfield  {journal} {\bibinfo
  {journal} {Phys. Rev. D}\ }\textbf {\bibinfo {volume} {92}},\ \bibinfo
  {pages} {016006} (\bibinfo {year} {2015})},\ \Eprint
  {http://arxiv.org/abs/1504.01308} {arXiv:1504.01308 [hep-ph]} \BibitemShut
  {NoStop}%
\bibitem [{\citenamefont {Villavicencio}\ \emph {et~al.}(2020)\citenamefont
  {Villavicencio}, \citenamefont {Dominguez},\ and\ \citenamefont
  {Loewe}}]{Villavicencio:2020fcz}%
  \BibitemOpen
  \bibfield  {author} {\bibinfo {author} {\bibfnamefont {C.}~\bibnamefont
  {Villavicencio}}, \bibinfo {author} {\bibfnamefont {C.~A.}\ \bibnamefont
  {Dominguez}}, \ and\ \bibinfo {author} {\bibfnamefont {M.}~\bibnamefont
  {Loewe}},\ }\href {\doibase 10.1088/1742-6596/1602/1/012027} {\bibfield
  {journal} {\bibinfo  {journal} {J. Phys. Conf. Ser.}\ }\textbf {\bibinfo
  {volume} {{1602}}},\ \bibinfo {pages} {{012027}} (\bibinfo {year} {2020})},\
  \Eprint {http://arxiv.org/abs/2007.05642} {arXiv:2007.05642 [hep-ph]}
  \BibitemShut {NoStop}%
\bibitem [{\citenamefont {Dominguez}(2018)}]{Dominguez:2018zzi}%
  \BibitemOpen
  \bibfield  {author} {\bibinfo {author} {\bibfnamefont {C.~A.}\ \bibnamefont
  {Dominguez}},\ }\href {\doibase 10.1007/978-3-319-97722-5} {\emph {\bibinfo
  {title} {{Quantum Chromodynamics Sum Rules}}}},\ SpringerBriefs in Physics\
  (\bibinfo  {publisher} {Springer International Publishing},\ \bibinfo
  {address} {Cham},\ \bibinfo {year} {2018})\BibitemShut {NoStop}%
\bibitem [{\citenamefont {Ioffe}(1981)}]{Ioffe:1981kw}%
  \BibitemOpen
  \bibfield  {author} {\bibinfo {author} {\bibfnamefont {B.}~\bibnamefont
  {Ioffe}},\ }\href {\doibase 10.1016/0550-3213(81)90259-5} {\bibfield
  {journal} {\bibinfo  {journal} {Nucl. Phys. B}\ }\textbf {\bibinfo {volume}
  {188}},\ \bibinfo {pages} {317} (\bibinfo {year} {1981})},\ \bibinfo {note}
  {[Erratum: Nucl.Phys.B 191, 591--592 (1981)]}\BibitemShut {NoStop}%
\bibitem [{\citenamefont {Chung}\ \emph {et~al.}(1981)\citenamefont {Chung},
  \citenamefont {Dosch}, \citenamefont {Kremer},\ and\ \citenamefont
  {Schall}}]{Chung:1981wm}%
  \BibitemOpen
  \bibfield  {author} {\bibinfo {author} {\bibfnamefont {Y.}~\bibnamefont
  {Chung}}, \bibinfo {author} {\bibfnamefont {H.~G.}\ \bibnamefont {Dosch}},
  \bibinfo {author} {\bibfnamefont {M.}~\bibnamefont {Kremer}}, \ and\ \bibinfo
  {author} {\bibfnamefont {D.}~\bibnamefont {Schall}},\ }\href {\doibase
  10.1016/0370-2693(81)91057-1} {\bibfield  {journal} {\bibinfo  {journal}
  {Phys. Lett. B}\ }\textbf {\bibinfo {volume} {102}},\ \bibinfo {pages} {175}
  (\bibinfo {year} {1981})}\BibitemShut {NoStop}%
\bibitem [{\citenamefont {Chung}\ \emph
  {et~al.}(1982{\natexlab{a}})\citenamefont {Chung}, \citenamefont {Dosch},
  \citenamefont {Kremer},\ and\ \citenamefont {Schall}}]{Chung:1981cc}%
  \BibitemOpen
  \bibfield  {author} {\bibinfo {author} {\bibfnamefont {Y.}~\bibnamefont
  {Chung}}, \bibinfo {author} {\bibfnamefont {H.~G.}\ \bibnamefont {Dosch}},
  \bibinfo {author} {\bibfnamefont {M.}~\bibnamefont {Kremer}}, \ and\ \bibinfo
  {author} {\bibfnamefont {D.}~\bibnamefont {Schall}},\ }\href {\doibase
  10.1016/0550-3213(82)90154-7} {\bibfield  {journal} {\bibinfo  {journal}
  {Nucl. Phys. B}\ }\textbf {\bibinfo {volume} {197}},\ \bibinfo {pages} {55}
  (\bibinfo {year} {1982}{\natexlab{a}})}\BibitemShut {NoStop}%
\bibitem [{\citenamefont {Chung}\ \emph
  {et~al.}(1982{\natexlab{b}})\citenamefont {Chung}, \citenamefont {Dosch},
  \citenamefont {Kremer},\ and\ \citenamefont {Schall}}]{Chung:1982rd}%
  \BibitemOpen
  \bibfield  {author} {\bibinfo {author} {\bibfnamefont {Y.}~\bibnamefont
  {Chung}}, \bibinfo {author} {\bibfnamefont {H.~G.}\ \bibnamefont {Dosch}},
  \bibinfo {author} {\bibfnamefont {M.}~\bibnamefont {Kremer}}, \ and\ \bibinfo
  {author} {\bibfnamefont {D.}~\bibnamefont {Schall}},\ }\href {\doibase
  10.1007/BF01614427} {\bibfield  {journal} {\bibinfo  {journal} {Z. Phys. C}\
  }\textbf {\bibinfo {volume} {15}},\ \bibinfo {pages} {367} (\bibinfo {year}
  {1982}{\natexlab{b}})}\BibitemShut {NoStop}%
\bibitem [{\citenamefont {Chung}\ \emph {et~al.}(1984)\citenamefont {Chung},
  \citenamefont {Dosch}, \citenamefont {Kremer},\ and\ \citenamefont
  {Schall}}]{Chung:1984gr}%
  \BibitemOpen
  \bibfield  {author} {\bibinfo {author} {\bibfnamefont {Y.}~\bibnamefont
  {Chung}}, \bibinfo {author} {\bibfnamefont {H.~G.}\ \bibnamefont {Dosch}},
  \bibinfo {author} {\bibfnamefont {M.}~\bibnamefont {Kremer}}, \ and\ \bibinfo
  {author} {\bibfnamefont {D.}~\bibnamefont {Schall}},\ }\href {\doibase
  10.1007/BF01557473} {\bibfield  {journal} {\bibinfo  {journal} {Z. Phys. C}\
  }\textbf {\bibinfo {volume} {25}},\ \bibinfo {pages} {151} (\bibinfo {year}
  {1984})}\BibitemShut {NoStop}%
\bibitem [{\citenamefont {Reinders}\ \emph {et~al.}(1985)\citenamefont
  {Reinders}, \citenamefont {Rubinstein},\ and\ \citenamefont
  {Yazaki}}]{Reinders:1984sr}%
  \BibitemOpen
  \bibfield  {author} {\bibinfo {author} {\bibfnamefont {L.}~\bibnamefont
  {Reinders}}, \bibinfo {author} {\bibfnamefont {H.}~\bibnamefont
  {Rubinstein}}, \ and\ \bibinfo {author} {\bibfnamefont {S.}~\bibnamefont
  {Yazaki}},\ }\href {\doibase 10.1016/0370-1573(85)90065-1} {\bibfield
  {journal} {\bibinfo  {journal} {Phys. Rept.}\ }\textbf {\bibinfo {volume}
  {127}},\ \bibinfo {pages} {1} (\bibinfo {year} {1985})}\BibitemShut {NoStop}%
\bibitem [{\citenamefont {Sadovnikova}\ \emph {et~al.}(2005)\citenamefont
  {Sadovnikova}, \citenamefont {Drukarev},\ and\ \citenamefont
  {Ryskin}}]{Sadovnikova:2005ye}%
  \BibitemOpen
  \bibfield  {author} {\bibinfo {author} {\bibfnamefont {V.}~\bibnamefont
  {Sadovnikova}}, \bibinfo {author} {\bibfnamefont {E.}~\bibnamefont
  {Drukarev}}, \ and\ \bibinfo {author} {\bibfnamefont {M.}~\bibnamefont
  {Ryskin}},\ }\href {\doibase 10.1103/PhysRevD.72.114015} {\bibfield
  {journal} {\bibinfo  {journal} {Phys. Rev. D}\ }\textbf {\bibinfo {volume}
  {72}},\ \bibinfo {pages} {114015} (\bibinfo {year} {2005})},\ \Eprint
  {http://arxiv.org/abs/hep-ph/0508240} {arXiv:hep-ph/0508240} \BibitemShut
  {NoStop}%
\bibitem [{\citenamefont {Nasrallah}\ and\ \citenamefont
  {Schilcher}(2014)}]{Nasrallah:2013ywh}%
  \BibitemOpen
  \bibfield  {author} {\bibinfo {author} {\bibfnamefont {N.~F.}\ \bibnamefont
  {Nasrallah}}\ and\ \bibinfo {author} {\bibfnamefont {K.}~\bibnamefont
  {Schilcher}},\ }\href {\doibase 10.1103/PhysRevC.89.045202} {\bibfield
  {journal} {\bibinfo  {journal} {Phys. Rev. C}\ }\textbf {\bibinfo {volume}
  {89}},\ \bibinfo {pages} {045202} (\bibinfo {year} {2014})},\ \bibinfo {note}
  {[Addendum: Phys.Rev.C 89, 059904 (2014)]},\ \Eprint
  {http://arxiv.org/abs/1310.6114} {arXiv:1310.6114 [hep-ph]} \BibitemShut
  {NoStop}%
\bibitem [{\citenamefont {Patrignani}\ \emph {et~al.}(2016)\citenamefont
  {Patrignani} \emph {et~al.}}]{Patrignani:2016xqp}%
  \BibitemOpen
  \bibfield  {author} {\bibinfo {author} {\bibfnamefont {C.}~\bibnamefont
  {Patrignani}} \emph {et~al.} (\bibinfo {collaboration} {Particle Data
  Group}),\ }\href {\doibase 10.1088/1674-1137/40/10/100001} {\bibfield
  {journal} {\bibinfo  {journal} {Chin. Phys. C}\ }\textbf {\bibinfo {volume}
  {40}},\ \bibinfo {pages} {100001} (\bibinfo {year} {2016})}\BibitemShut
  {NoStop}%
\bibitem [{\citenamefont {Kamikado}\ and\ \citenamefont
  {Kanazawa}(2014)}]{Kamikado:2013pya}%
  \BibitemOpen
  \bibfield  {author} {\bibinfo {author} {\bibfnamefont {K.}~\bibnamefont
  {Kamikado}}\ and\ \bibinfo {author} {\bibfnamefont {T.}~\bibnamefont
  {Kanazawa}},\ }\href {\doibase 10.1007/JHEP03(2014)009} {\bibfield  {journal}
  {\bibinfo  {journal} {JHEP}\ }\textbf {\bibinfo {volume} {03}},\ \bibinfo
  {pages} {009} (\bibinfo {year} {2014})},\ \Eprint
  {http://arxiv.org/abs/1312.3124} {arXiv:1312.3124 [hep-ph]} \BibitemShut
  {NoStop}%
\bibitem [{\citenamefont {Pisarski}\ and\ \citenamefont
  {Tytgat}(1996)}]{Pisarski:1996mt}%
  \BibitemOpen
  \bibfield  {author} {\bibinfo {author} {\bibfnamefont {R.~D.}\ \bibnamefont
  {Pisarski}}\ and\ \bibinfo {author} {\bibfnamefont {M.}~\bibnamefont
  {Tytgat}},\ }\href {\doibase 10.1103/PhysRevD.54.R2989} {\bibfield  {journal}
  {\bibinfo  {journal} {Phys. Rev. D}\ }\textbf {\bibinfo {volume} {54}},\
  \bibinfo {pages} {2989} (\bibinfo {year} {1996})},\ \Eprint
  {http://arxiv.org/abs/hep-ph/9604404} {arXiv:hep-ph/9604404} \BibitemShut
  {NoStop}%
\bibitem [{\citenamefont {Son}\ and\ \citenamefont
  {Stephanov}(2000)}]{Son:1999cm}%
  \BibitemOpen
  \bibfield  {author} {\bibinfo {author} {\bibfnamefont {D.}~\bibnamefont
  {Son}}\ and\ \bibinfo {author} {\bibfnamefont {M.~A.}\ \bibnamefont
  {Stephanov}},\ }\href {\doibase 10.1103/PhysRevD.61.074012} {\bibfield
  {journal} {\bibinfo  {journal} {Phys. Rev. D}\ }\textbf {\bibinfo {volume}
  {61}},\ \bibinfo {pages} {074012} (\bibinfo {year} {2000})},\ \Eprint
  {http://arxiv.org/abs/hep-ph/9910491} {arXiv:hep-ph/9910491} \BibitemShut
  {NoStop}%
\bibitem [{\citenamefont {Bali}\ \emph {et~al.}(2012)\citenamefont {Bali},
  \citenamefont {Bruckmann}, \citenamefont {Endrodi}, \citenamefont {Fodor},
  \citenamefont {Katz},\ and\ \citenamefont {Schafer}}]{Bali:2012zg}%
  \BibitemOpen
  \bibfield  {author} {\bibinfo {author} {\bibfnamefont {G.}~\bibnamefont
  {Bali}}, \bibinfo {author} {\bibfnamefont {F.}~\bibnamefont {Bruckmann}},
  \bibinfo {author} {\bibfnamefont {G.}~\bibnamefont {Endrodi}}, \bibinfo
  {author} {\bibfnamefont {Z.}~\bibnamefont {Fodor}}, \bibinfo {author}
  {\bibfnamefont {S.}~\bibnamefont {Katz}}, \ and\ \bibinfo {author}
  {\bibfnamefont {A.}~\bibnamefont {Schafer}},\ }\href {\doibase
  10.1103/PhysRevD.86.071502} {\bibfield  {journal} {\bibinfo  {journal} {Phys.
  Rev. D}\ }\textbf {\bibinfo {volume} {86}},\ \bibinfo {pages} {071502}
  (\bibinfo {year} {2012})},\ \Eprint {http://arxiv.org/abs/1206.4205}
  {arXiv:1206.4205 [hep-lat]} \BibitemShut {NoStop}%
\bibitem [{\citenamefont {Tanabashi}\ \emph {et~al.}(2018)\citenamefont
  {Tanabashi} \emph {et~al.}}]{Tanabashi:2018oca}%
  \BibitemOpen
  \bibfield  {author} {\bibinfo {author} {\bibfnamefont {M.}~\bibnamefont
  {Tanabashi}} \emph {et~al.} (\bibinfo {collaboration} {Particle Data
  Group}),\ }\href {\doibase 10.1103/PhysRevD.98.030001} {\bibfield  {journal}
  {\bibinfo  {journal} {Phys. Rev. D}\ }\textbf {\bibinfo {volume} {98}},\
  \bibinfo {pages} {030001} (\bibinfo {year} {2018})}\BibitemShut {NoStop}%
\bibitem [{\citenamefont {Dominguez}\ and\ \citenamefont
  {Loewe}(1993)}]{Dominguez:1992vg}%
  \BibitemOpen
  \bibfield  {author} {\bibinfo {author} {\bibfnamefont {C.~A.}\ \bibnamefont
  {Dominguez}}\ and\ \bibinfo {author} {\bibfnamefont {M.}~\bibnamefont
  {Loewe}},\ }\href {\doibase 10.1007/BF01560345} {\bibfield  {journal}
  {\bibinfo  {journal} {Z. Phys. C}\ }\textbf {\bibinfo {volume} {58}},\
  \bibinfo {pages} {273} (\bibinfo {year} {1993})}\BibitemShut {NoStop}%
\bibitem [{\citenamefont {Bali}\ \emph {et~al.}(2020)\citenamefont {Bali},
  \citenamefont {Endr{\"o}di},\ and\ \citenamefont {Piemonte}}]{Bali:2020bcn}%
  \BibitemOpen
  \bibfield  {author} {\bibinfo {author} {\bibfnamefont {G.~S.}\ \bibnamefont
  {Bali}}, \bibinfo {author} {\bibfnamefont {G.}~\bibnamefont {Endr{\"o}di}}, \
  and\ \bibinfo {author} {\bibfnamefont {S.}~\bibnamefont {Piemonte}},\ }\href
  {\doibase 10.1007/JHEP07(2020)183} {\bibfield  {journal} {\bibinfo  {journal}
  {JHEP}\ }\textbf {\bibinfo {volume} {07}},\ \bibinfo {pages} {183} (\bibinfo
  {year} {2020})},\ \Eprint {http://arxiv.org/abs/2004.08778} {arXiv:2004.08778
  [hep-lat]} \BibitemShut {NoStop}%
\end{thebibliography}%

\end{document}